\documentclass[apl,aps,twocolumn,superscriptaddress]{revtex4-1}

\usepackage{amsmath}
\usepackage{graphicx} 
\usepackage{dcolumn} 
\usepackage{bm} 
\usepackage{color} 
\usepackage{epstopdf}
\usepackage{amssymb}
\usepackage{verbatim}
\usepackage[colorlinks=true, pdfstartview=FitV, linkcolor=blue, citecolor=blue, urlcolor=blue]{hyperref} 
\usepackage{epstopdf}

\begin{document}

\newcommand{\addMD}[1]{\textcolor{magenta}{#1}}

\newcommand{\ee}[1]{\times 10^{#1}}
\newcommand{\mr}[1]{\mathrm{#1}}
\newcommand{\unit}[1]{\,\mathrm{#1}}
\newcommand{\um}{\,\mu{\rm m}}
\newcommand{\us}{\,\mu{\rm s}}
\newcommand{\uT}{\,\mu{\rm T}}
\newcommand{\kT}{k_{\rm B}T}
\newcommand{\kB}{k_{\rm B}}
\newcommand{\muB}{\mu_{\rm B}}
\newcommand{\rtHz}{\sqrt{\mr{Hz}}}
\newcommand{\degree}{^\circ}

\newcommand{\yn}{\gamma_n}
\newcommand{\tc}{t_c}
\newcommand{\fc}{f_c}
\newcommand{\fRF}{f_\mr{RF}}
\newcommand{\Dfdev}{\Delta f_\mr{dev}}
\newcommand{\fcenter}{f_\mr{center}}
\newcommand{\fL}{f_\mr{L}}
\newcommand{\Tp}{T_\mr{p}}
\newcommand{\tm}{\tau_\mr{m}}
\newcommand{\Qdamped}{Q_\mr{damped}}
\newcommand{\ave}[1]{\left\langle #1\right\rangle}

\title{A parametric symmetry breaking transducer}%
\author{Alexander Eichler}
\affiliation{Institute for Solid State Physics, ETH Zurich, 8093 Zurich, Switzerland}
\author{Toni L. Heugel}
\affiliation{Institute for Theoretical Physics, ETH Zurich, 8093 Zurich, Switzerland}
\author{Anina Leuch}
\affiliation{Institute for Solid State Physics, ETH Zurich, 8093 Zurich, Switzerland}
\affiliation{Institute for Theoretical Physics, ETH Zurich, 8093 Zurich, Switzerland}
\author{Christian L. Degen}
\affiliation{Institute for Solid State Physics, ETH Zurich, 8093 Zurich, Switzerland}
\author{R. Chitra}
\affiliation{Institute for Theoretical Physics, ETH Zurich, 8093 Zurich, Switzerland}
\author{Oded Zilberberg}
\affiliation{Institute for Theoretical Physics, ETH Zurich, 8093 Zurich, Switzerland}
\date{\today}
\begin{abstract}
Force detectors rely on resonators to transduce forces into a readable signal. 
Usually  these resonators operate in the linear regime and their signal appears amidst a competing background comprising  thermal or quantum fluctuations as well as readout noise. Here, we demonstrate that a parametric symmetry breaking transduction leads to a novel and robust nonlinear force detection in the presence of noise. The force signal  is encoded in the frequency at which the system jumps between two phase states which are inherently protected against phase noise. Consequently, the transduction effectively decouples from readout noise channels. For a controlled demonstration of the method, we experiment with a macroscopic doubly-clamped string. Our method provides a promising new paradigm for high-precision force detection.
\end{abstract}
\maketitle


Resonators are widely used for the detection and amplification of oscillating signals. In its most basic and ubiquitous form, resonant detection measures the amplitude of oscillation in response to a signal. Examples of resonator-based sensors range from radar antennas~\cite{Skolnik_2000} and nuclear magnetic resonance~\cite{Rabi_1938} to optical antennas~\cite{Novotny_2011}, to gravitational wave detection~\cite{Weber_1960,Abbott_2016}, and to nanomechanical force transducers~\cite{Binnig_1986,Rugar_1990,Mamin_2001,Arlett_2006}. An attractive feature of resonant detection is the possibility of phase-sensitive signal transduction, which can be used to reject unwanted or incoherent signal sources in a lock-in type measurement. In magnetic resonance force microscopy (MRFM), for instance, a small magnetic force acting on a force transducer (a cantilever) is modulated at the transducer's resonance frequency and drives coherent oscillations~\cite{Sidles_1991,Rugar_2004,Poggio_2010}. The controlled phase and frequency of the force modulation allows to distinguish weak force signals against an overwhelming noise background.

The sensitivity of a detector is limited by intrinsic fluctuations and by readout noise, both of which can obscure the true response to the force signal. Intrinsic fluctuations include amplitude and phase noise of the resonator vibrations and can stem from many sources. In the particular case of a  classical mechanical force transducer, the lowest limit of intrinsic fluctuations is set by the white thermomechanical force noise.  This threshold  can be decreased by designing resonators with small masses and high mechanical quality factors~\cite{Moser_2013,Tsaturyan_2017,Heritier_2018}. Readout noise, on the other hand, is added in the signal amplification process and is typically more pronounced when the resonator vibrations are small. Therefore, as force sensors are scaled down, they usually experience a \textit{decrease} of intrinsic fluctuations as well as an \textit{increase} of readout noise. This tradeoff establishes a lower boundary for the forces that can be detected. Pushing this boundary is vital for all sensing techniques.

Standard parametric amplification can reduce readout noise by amplifying the resonator's motion using purely reactive components. Examples of its application include (i) varactor amplifiers used for radio signals~\cite{Heffner_1958,Penfield_1962}, (ii) superconducting parametric amplifiers that have demonstrated readout noise close to that imposed by the laws of quantum mechanics~\cite{Kuzmin_1983,Yurke_1989,Roy_2016}, (iii) squeezed mechanical vibrations~\cite{Rugar_1991,Karabalin_2011,Szorkovszky_2011,Poot_2014,Mahboob_2014}, as well as proposals for improved sensitivity of gravitational waves detection~\cite{Caves_1981}. Such techniques, however, are bound to operate in a regime of relatively small vibrations, i.e., well below the parametric instability threshold~\cite{Lifshitz_Cross}. 

\begin{figure}
\includegraphics[width=\columnwidth]{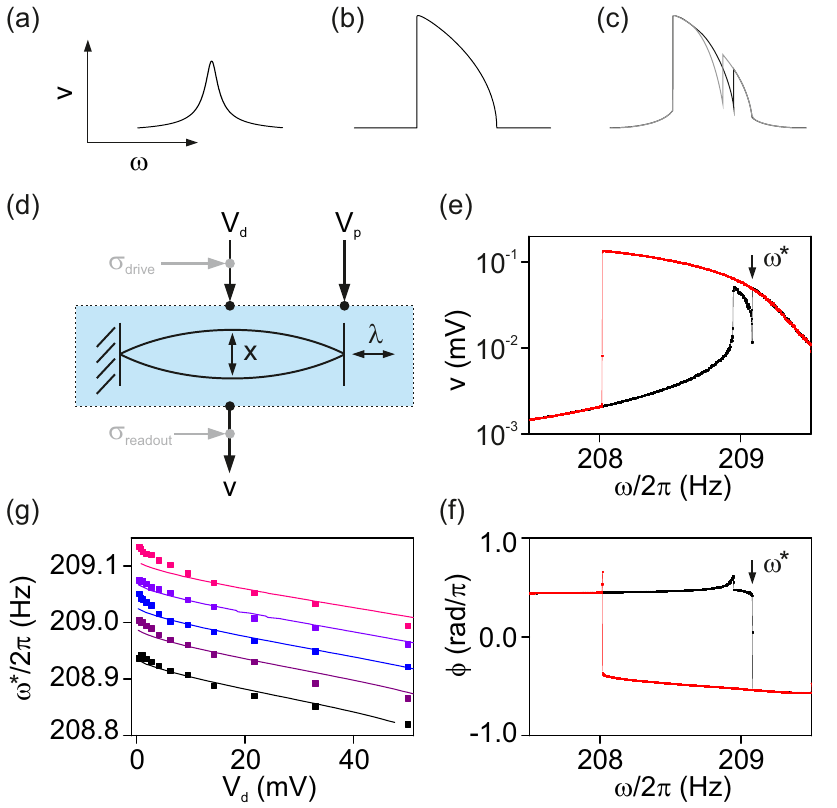}
\caption{\label{fig1} (a-c) Schematic working principle of the PSBT. Response of the resonator to a small force (a) or to a large parametric drive (b), individually. When combined, the two driving sources produce a complex response that has a characteristic second jump (c). Grey and black lines demonstrate how the jump frequency changes as a function of the applied force. (d) Sketch of the setup showing input voltages $V_d$ and $V_p$, as well as the measured output voltage $v$ (which corresponds to a mechanical vibration $x$). Added voltage noise sources are indicated by their standard deviations $\sigma_{\rm drive}$ and $\sigma_{\rm readout}$. (e-f) Amplitude and phase of a frequency sweep with parametric drive and external drive applied simultaneously, with $V_p = 0.25$\,V and $V_d = 0.1$\,V. Red and black correspond to sweeps with decreasing and increasing frequencies, respectively. The arrow marks the frequency $\omega^{*}$ at which the PSBT jump occurs. (g) Position of frequency jump, $\omega^*$, as a function of applied voltage, $V_d$. The curves were offset for better visibility and correspond to $V_p=0.21$\,V -- $0.25$\,V in steps of $0.01$\,V (bottom to top). Lines denote theory fits using $\alpha = -6.7 \times 10^6$\,V$^{-2}$s$^{-2}$ as the only fitting parameter. Data sets are offset for visibility and slow frequency drifts occurring between sets were compensated.}
\end{figure}

In this work, we experimentally demonstrate for the first time  a complementary approach for sensitive force detection. In contrast to standard parametric amplification,   this method operates beyond the instability threshold.  It employs a parametrically driven, nonlinear resonator where the presence of a small external force leads to a distinct double-hysteresis pattern in a frequency sweep~\cite{Leuch_2016}. This double hysteresis allows measuring the applied force via the parametric symmetry breaking transducer (PSBT) method, see Figs.~\ref{fig1}(a)-(c). Importantly, even though the resonator vibrations are inherently nonlinear, we show that the transducer has an approximately linear gain characteristic.  In comparison with linear transducers,  the PSBT performance degrades  faster  in the presence of large intrinsic fluctuations.  However, it is highly insensitive to readout noise, which makes it, for example, promising for applications with nanomechanical force sensors. 

The PSBT can be realized with any system that fulfills the following equation
\begin{multline} \label{eq:Newton_equation}
\ddot{v} + \omega_0^2 \left[ 1 - \lambda \cos \left(  2 \omega t  \right) \right] v + 
\Gamma \dot{v} + \alpha v^3  \\= C\cdot V_d \cos \left( \omega t  + \phi \right)  + C\cdot\xi(t)\,.
\end{multline}
Here, we have chosen a representation in electrical units to emphasize the generality of the physics involved, i.e., $v = v(t)$ is the measured voltage that is roughly proportional to the resonator displacement. Dots denote differentiation with respect to time $t$, $\omega$ is a frequency close to the angular eigenfrequency $\omega_0$ of the resonator, $\Gamma$ is a linear damping constant, and $\alpha$ represents a nonlinear (Duffing) spring constant. $V_d$ is the amplitude of an applied external driving voltage that is proportional to a force applied to the resonator with phase $\phi$, and $C$ is a gain factor~\cite{supmat}. $\xi(t)$ is an additive intrinsic fluctuating drive with standard deviation $\sigma_{\rm drive}$. In addition to the external drive, we also modulate the resonance frequency at a rate $2\omega$ and with a modulation depth $\lambda$, which we refer to as `parametric drive' and which we control with a voltage signal of amplitude $V_p$. Beyond a threshold value $V_{th}\propto\lambda_{th} = 2\Gamma/\omega_0$, this excitation leads to large and stable oscillations of the resonator. In our system, there is an additional nonlinear damping term in Eq.~\eqref{eq:Newton_equation}, $\eta v^2 \dot{v}$, that has negligible influence on the PSBT performance.

Our experimental demonstration is based on a macroscopic doubly clamped string that vibrates mechanically in accord with Eq. \eqref{eq:Newton_equation}, see Fig.~\ref{fig1}(d)~\cite{Leuch_2016} (see \cite{supmat} for a derivation of Eq. \eqref{eq:Newton_equation} for a mechanical resonator). We characterize the resonator at both small and large vibration $v \propto x$. With a small external force applied and with $\lambda = 0$, the resonator behaves linearly [Fig.~\ref{fig1}(a)], which allows us to extract $\omega_0/(2 \pi) \sim  208.8$\,Hz, $\Gamma = \omega_0/Q$ with quality factor $Q = 2150$, and $C = 430$\,s$^{-2}$. To fit the nonlinear constants $\alpha$ and $\eta$, we set $V_d = 0$ and drive the system to large amplitudes with $V_p > V_{th} = 80$\,mV. From a fit to the nonlinear response [Fig.~\ref{fig1}(b)] we obtain $\alpha / \eta = 1875$\,s$^{-1}$~\cite{supmat}.


\begin{figure}
\includegraphics[width=\columnwidth]{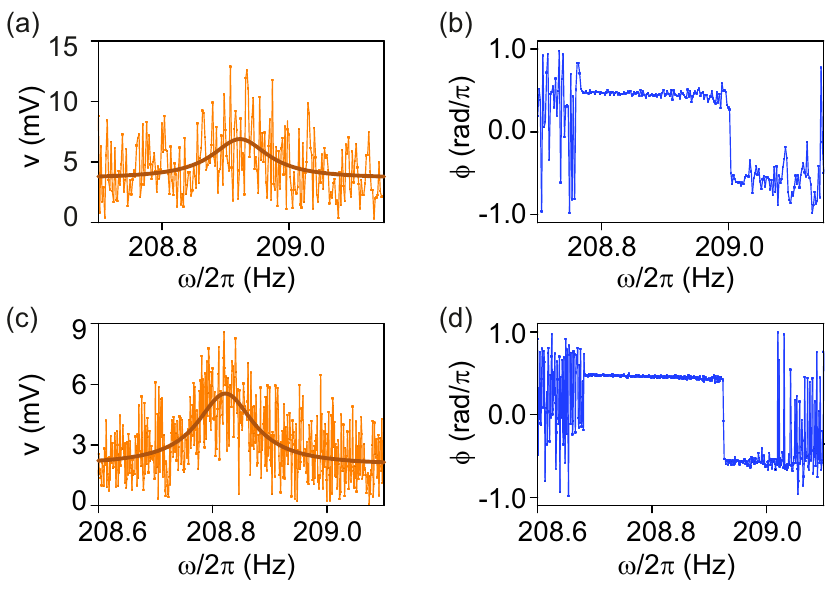}
\caption{\label{fig2} (a) Linear sensor ($V_p = 0$) with artificial readout noise (standard deviation $\sigma_{\rm readout}=1.8$\,mV within the measurement bandwidth). A voltage with amplitude $V_d = 10$\,mV was swept in frequency and the response detected with a lock-in amplifier. Total sweep time was $1920$\,s. (b) The same sweep performed in PSBT mode ($V_p = 0.25$\,V). Here we detect the phase rather than the amplitude. (c) Linear sensor with artificial intrinsic noise power spectral density $S_{\rm drive}=2.96 \times 10^{-4}$\,V$^2$/Hz and $V_d = 10$\,mV. Total sweep time was $4080$\,s. (d) The same sweep performed in PSBT mode ($V_p = 0.25$\,V). Solid lines in (a) and (c) are Lorentzian fits.}
\end{figure}

To perform force measurements, we exploit the double-hysteresis pattern that emerges when parametric and external drives are applied simultaneously [Fig.~\ref{fig1}(e)]. The underlying physics is governed by a symmetry breaking in the parametric phase states~\cite{Rhoads_2010,Papariello_2016,Leuch_2016}. The second jump of the upsweep at $\omega^{*}$ is a direct consequence of the interplay between the two drives. In the absence of noise, the jump frequency is expected to depend approximately linearly on $V_d$ for a range of forces. In the following, we shall focus on the accompanying phase jump at $\omega^*$, which corresponds roughly to $\pi$ radians even when the jump is small in amplitude [Fig.~\ref{fig1}(f)]. The phase jump is the most convenient experimental signature for our force detection method.  

In Fig.~\ref{fig1}(g), we experimentally demonstrate the relationship between $\omega^*$ and $V_d$ for various values of the parametric modulation depth. The corresponding theoretical results are obtained by studying the  time-averaged slow dynamics of the system and the jump frequency $\omega^*$  is derived using a bifurcation analysis of the equations of motion~\cite{supmat}. The almost-linear dependence of $\omega^{*}$ on $V_d$ indicates that usage of the calibrated force sensor is straightforward in spite of the complex nonlinear physics involved. 

We now evaluate the performance of our method in the presence of noise. Since our resonator operates far above any natural noise levels, we artificially add white voltage noise either in the form of intrinsic fluctuations $\xi(t)$ or as a fluctuating component of the measured voltage $v$, with standard deviations $\sigma_{\rm drive}$ and $\sigma_{\rm readout}$, respectively [see Fig.~\ref{fig1}(d)]. For comparison, we first use the resonator as a simple linear transducer without parametric drive. In Figs.~\ref{fig2}(a) and (c), we show that the resulting amplitude signal in the presence of either of the noise channels is almost entirely obscured by the fluctuations. Next, we repeat the sweep with added parametric drive to operate the resonator as a PSBT. Even though the fluctuations  in the phase are noticeable in the signal, the phase jump stands out clearly in Figs.~\ref{fig2}(b) and (d). 

\begin{figure}
\includegraphics[width=\columnwidth]{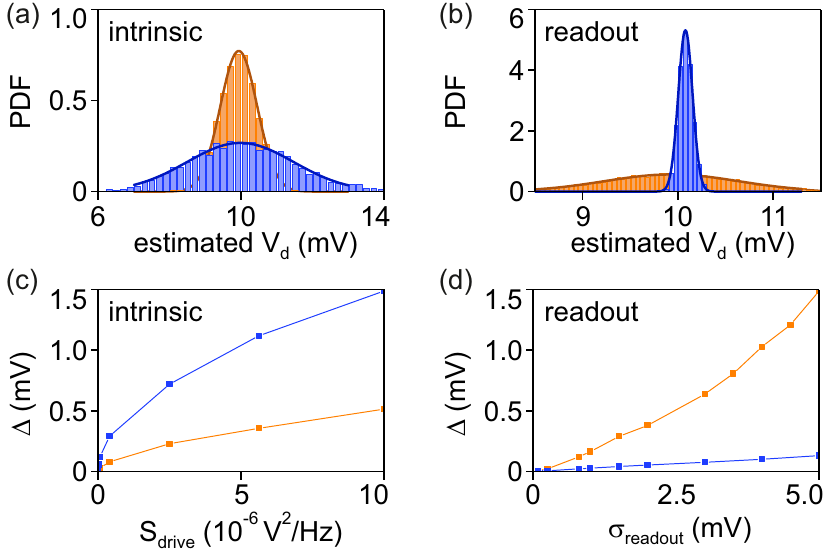}
\caption{\label{fig3} (a) and (b) probability density function (PDF) of the driving voltage $V_d$ estimated from simulated sweeps in the presence of intrinsic noise and readout noise, respectively. Orange bars denote results obtained with the linear sensor, blue bars are for the PSBT sensor (note corresponding colors in Fig.~\ref{fig2}). Solid lines correspond to Gaussian fits that allow extracting the standard error ($\Delta$). All resonator parameters were matched to the experimental device. The intrinsic noise in (a) has a power spectral density of $S_{\rm drive}=10^{-5}$\,V$^2$/Hz and the readout noise in (b) has a standard deviation of $3$\,mV. In each case, we simulated $3000$ sweeps with intrinsic noise and $10,000$ sweeps with readout noise. (c) and (d) standard error of estimated $V_d$ as a function of the standard deviation of applied intrinsic and readout noise, respectively.}
\end{figure}

Comparing the linear transducer and the PSBT, we have shown that the latter exhibits a surprising robustness to both intrinsic and readout noise. However, this does not yet characterize the precision of the PSBT in estimating the amplitude of the external signal $V_d$ from the jump frequency $\omega^*$, namely, we need to know the variance of the PSBT estimation. To this end, we numerically simulate repeated  measurements of $V_d$ using both the linear and PSBT methods. The simulations enable us to obtain large statistics of the detection performance in the presence of controllable and independent noise channels. In Figs.~\ref{fig3}(a) and (b), exemplary histograms of the estimated signal are presented for readout and intrinsic noise channels, respectively. The PSBT histograms exhibit an almost-Gaussian distribution  whose standard deviation  quantifies  a  standard error $\Delta$~\cite{supmat} for the force measurement. 

In Figs.~\ref{fig3}(c) and (d), we show  how noise influences $\Delta$. We systematically observe that  the impact of intrinsic noise on the PSBT is larger than its effect on the corresponding  linear transducer. The intrinsic fluctuations increase the chance for the PSBT to flip prematurely, which translates into frequency noise in the estimation. However, for readout noise, the situation is manifestly opposite and the PSBT significantly outperforms the linear transducer. This is a direct consequence of the fact that the PSBT signal is encoded in the phase of the oscillation, while the phase noise is reduced by driving the oscillator to a large amplitude. The PSBT, thus, effectively decouples from the readout noise channel. Our analysis indicates that the PSBT will have a better signal-to-noise ratio in situations where the detection is limited by readout noise.

Finally, we would like to discuss some limitations of the PSBT scheme: (i) the PSBT relies on a joint sweep of the frequencies of both external and parametric drives. This implies that the measured force can be modulated at a desired frequency and with a controlled phase, similar to MRFM~\cite{Sidles_1991}; (ii) the dynamic range, i.e. the range of forces that can be measured, depends on the parametric drive. When the resonator amplitude in response to the measured force becomes comparable to that of the parametric oscillation, the double hysteresis is replaced by a qualitatively different behavior and the PSBT scheme breaks down. In our experiment and for $V_d = 0.25$\,V, this resulted in an upper limit of $V_d \approx 0.15$\,V; (iii) the PSBT method is sensitive to frequency noise. Fluctuations of $\omega_0$ will lead to shifts in $\omega^*$ and distort the estimation of the measured force; (iv) the bandwidth (i.e. the repetition rate) of force measurements with our method is given by the sweeping speed, and therefore ultimately by the resonator's quality factor.


We have demonstrated that the PSBT has several attractive features that set it apart from a linear force transducer. The PSBT makes use of parametric phase states, which are intrinsically protected against amplitude and phase noise. Since the measured force is extracted from a frequency as opposed to   amplitude, the PSBT can measure small forces  even while operating at  relatively large oscillation amplitudes.  This feature  makes the PSBT highly tolerant to readout noise similar to  frequency-modulated, feedback-driven oscillators~\cite{Albrecht_1991}. However, in contrast to the slow frequency modulation rate used in the latter, our method can detect forces at frequencies close to the eigenfrequency of the transducer itself.  We believe that the PSBT is promising for force detection experiments with nanomechanical resonators such as carbon nanotubes or graphene, as well as for the detection of electrical signals with Josephson parametric resonators. Further work will focus on the performance of PSBT sensors in the quantum realm.

\begin{acknowledgments}
We acknowledge fruitful discussions with L.~Papariello and technical support from C. Keck, P. M\"arki, M. Baer and the mechanical workshop team of the Department of Physics at ETH Zurich. This work received financial support from the Swiss National Science Foundation (CRSII5\_177198/1).
\end{acknowledgments}

\clearpage

\large
\begin{center}

\textbf{Supplementary Material for: A parametric symmetry breaking transducer}

\end{center}

\small


\renewcommand{\thefigure}{S\arabic{figure}}
\renewcommand{\theequation}{S\arabic{equation}}
\renewcommand{\bibnumfmt}[1]{[S#1]}
\renewcommand{\citenumfont}[1]{S#1}
\setcounter{equation}{0}
\setcounter{figure}{0}
\renewcommand{\figurename}{\textbf{Supplementary Figure}}


\section{Derivation of the equation of motion in units of volts}
In our work, we analyze the behavior of a mechanical system that is controlled by, and observed via, AC voltage signals. It is therefore natural to describe the dynamics of the system directly in units of volts. This is also in line with the fact that our analysis applies to a general sensor prototype, and not specifically to a mechanical resonator. We start with the equation of motion for a nonlinear resonator:
\begin{multline} \label{eq:S1}
\ddot{x} + \omega_0^2 \left[ 1 - \lambda \cos \left(  2 \omega t  \right) \right] x + \Gamma \dot{x} + \frac{\tilde{\alpha}}{M} x^3 + \frac{\tilde{\eta}}{M} x^2 \dot{x} \\= \frac{F_0}{M} \cos \left( \omega t  + \phi \right)\,,
\end{multline}
where $x$ is the resonator displacement, $\tilde{\alpha}$ is a mechanical nonlinear spring constant in units of kgm$^{-2}$s$^{-2}$, $\tilde{\eta}$ is the nonlinear damping coefficient in units of kgm$^{-2}$s$^{-1}$, $M$ is the effective mass of the resonator, and $F_0$ the driving force in units of N (see main text for definition of $\omega_0$, $\omega$, $\lambda$, $\Gamma$, $t$ and $\phi$). In order to obtain a description in terms of voltage instead of displacement and force, we introduce a proportionality factor $c$ between displacement and measured voltage, $x \times c = v$, and $C_1$ between driving voltage and mechanical force, $F_0 = V_{d} \times C_1 M$, and obtain
\begin{multline}
\ddot{v} + \omega_0^2 \left[ 1 - \lambda \cos \left(  2 \omega t  \right) \right] v + \Gamma \dot{v} + \alpha v^3 + \eta v^2 \dot{v} \\= V_d \cos \left( \omega t  + \phi \right) \times C_1 \times c\,,
\end{multline}
where $\alpha = \tilde{\alpha} / c^2 M$ and $\eta = \tilde{\eta} / c^2 M$. The numerical values of $\alpha$ and $\eta$ therefore depend on the choice of $c$. Without loss of generality, we can set $c = 1$\,V/m and $C=C_1 c$ to arrive at eq. (1) of the main text. Please note that the nonlinear damping term $\eta$ is unimportant for the presented study and is therefore omitted in the main text.

\begin{figure}[t!]
\includegraphics[width=\columnwidth]{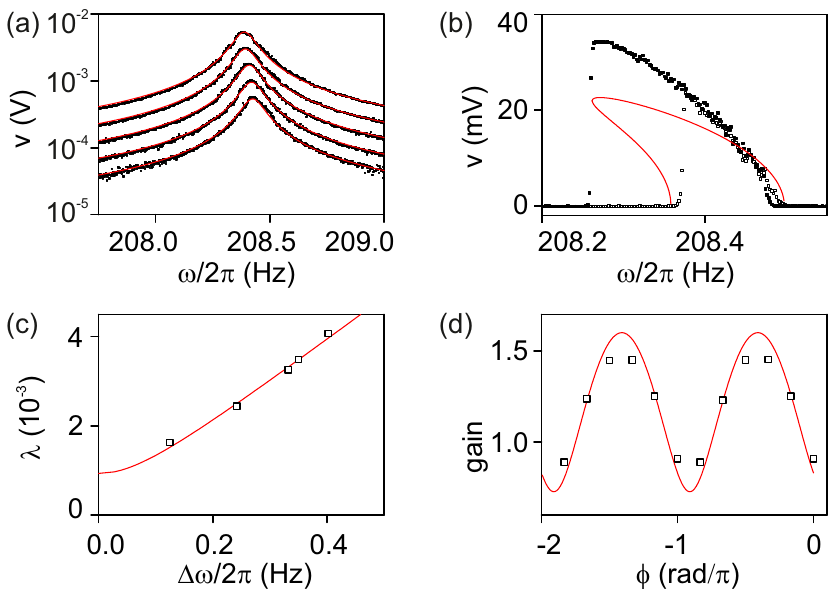}
\caption{\label{fig:figureS1} (a) Linear resonator regime without parametric drive. The driving voltage was $1$, $1.78$, $3.2$, $5.6$ and $10$\,mV from bottom to top. (b) Nonlinear regime with parametric drive only. Open and full dots correspond to sweeps with increasing and decreasing frequency, respectively. A red line shows the model calculation with the parameter values stated in the main text and with $V_p = 140$\,mV. (c) Width of the parametric instability region as a function of modulation depth $\lambda$. The resonator becomes unstable beyond $\lambda_{th} = 2/Q$. The experimental data are shown as dots and were used to extract the proportionality between $V_p$ and $\lambda$. From the fit, we find a threshold voltage of $V_{th} = 80$\,mV. (d) Subthreshold parametric amplification as a function of phase $\phi$. Data points correspond to the maximum amplitude of a sweep with constant phase relative to that of a sweep with no parametric amplification. The points were shifted with a global phase $\Delta \phi = -1.07$\,rad to fit the theory (red line).}
\end{figure}

\section{Experimental system and basic characterization}
The experimental setup has been described in detail in a previous publication \cite{Leuch_2016}. The resonating element is a guitar string that is rigidly clamped on one end while the other end can be subjected to small displacements along the string axis. These displacements change the tension along the string and modulate the spring constant in the typical sense of a parametric drive \cite{Lifshitz_Cross,Papariello_2016}. In addition, the string can be actuated by an external force through coils that apply a magnetic field gradient to the magnetized string. The string vibrations are detected inductively using a commercial Seymour Duncan Humbucker and a lock-in amplifier (Zurich Instruments HF2LI). The string used in the present experiment is the G string of a EXL120 D'Addario set. It has a diameter of $0.406$\,mm and was suspended over a length of $0.36$\,m, from which we calculate an effective mass of $M = 1.9 \times 10^{-4}$\,kg.

We performed basic characterization of the resonator to get numerical values for its parameters. Figure \ref{fig:figureS1}(a) shows sweeps with small external forces applied without parametric modulation ($\lambda = 0$). The linear response of the resonator allows us to extract $\omega_0$ (between $208$ and $209$\,Hz depending on temperature), mechanical quality factor $Q = 2150$, $C = 430$\,s$^{-2}$, and a direct inductive background between driving and pickup coils of $5 \times 10^{-4}$\,V/V$_d$. This background is only important for calibration purposes and does not affect the operation of the PSBT. When driven with strong parametric modulation in the absence of an external drive ($V_d = 0$), the resonator reaches high amplitudes and is in the nonlinear regime. From Fig. \ref{fig:figureS1}(b), we find that the amplitude as a function of frequency cannot be precisely described by the usual model \cite{Lifshitz_Cross}, which we ascribe to a nonlinearity in the inductive detection method. Indeed, in this regime, the string vibration amplitude is of the same order of magnitude as its separation to the pickup coil, which can lead to nonlinear transduction. Note that the value of our Duffing nonlinearity, $\alpha = -6.7 \times 10^6$\,V$^{-2}$s$^{-2}$, was extracted from the jump frequency $\omega^*$ in Fig. 1g of the main text, which is not affected by the readout nonlinearity. We can accurately assign $\alpha/\eta = -1.88\times 10^{4}$\,s$^{-1}$ from the frequency of the left jump at about $208.2$\,Hz. Figure \ref{fig:figureS1}(c) shows the width of the parametric instability region, the so-called `Arnold tongue', as a function of modulation depth $\lambda$. Open squares correspond to the measured data, a red line to the theory. We used a conversion factor between $\lambda$ and $V_p$ as fitting parameter, which yielded a parametric threshold voltage of $V_{th} = 80$\,mV. Finally, we calibrated the phase offset $\Delta\phi$ between the parametric drive and the external force (due to inductive elements etc.) from the phase dependence of subthreshold parametric amplification \cite{Lifshitz_Cross,Papariello_2016}. We get $\Delta \phi = -1.07$\,rad, which adds to the set phase $\Delta \phi = -2.36$\,rad used in the experiments.

\section{Theoretical estimation of the jump frequency}

The jump frequency at which the second hysteresis occurs can be directly estimated using a bifurcation analysis of the resonator's equation of motion. The equation of motion [Eq.~\eqref{eq:S1}] is typically studied using the averaging method \cite{Papariello_2016, Guckenheimer_Holmes}, which replaces the full time-dependent equation by time-independent averaged equations of motion. We rewrite, Eq.~\eqref{eq:S1} in terms of dimensionless variables, $\tau = \omega_0 t$ and $z= \sqrt{\alpha/(m \omega_0^2)}x$:
\begin{equation} \label{eq:S3}
		\resizebox{0.88\hsize}{!}{$\ddot{z} + \bar{\gamma} \dot{z} + z^3 + \bar{\eta} z^2 \dot{z} + (1-\lambda \cos( 2\Omega \tau)) z = \bar{F}_0 \cos(\Omega \tau + \phi)$,}
\end{equation}
where the dimensionless parameters are defined as $\bar{\gamma} \equiv \gamma/(m \omega_0)$, $\bar{\eta} \equiv \eta \omega_0/|\alpha|$, $\Omega \equiv \omega/\omega_0$, and $\bar{F}_0 \equiv (F_0/\omega_0^3) \sqrt{|\alpha| /m^3}$. As shown in Ref.~\cite{Papariello_2016}, for driving around the first instability lobe, with $\omega \approx \omega_0$ and using the van der Pol transformation to variables $U$ and $V$, the full dynamics of the parametric resonator can be described by the slow-flow variables $u = \bar{U}$ and $v = \bar{V}$, which correspond to $U$ and $V$ averaged over one period of the parametric drive. The corresponding slow-flow equations are:
\begin{widetext}
\begin{align}
\dot{u} &= f_1(u,v) = - \frac{1}{2 \Omega} \left[ \bar{\gamma} \Omega  u + v \left(\sigma + \frac{\lambda}{2} \right) + \frac{3}{4} \left( u^2 + v^2 \right) v + \Omega \frac{\bar{\eta}}{4}  \left( u^2 + v^2 \right) u - \bar{F}_0 \sin(\phi) \right] \label{eq:S4}\,, \\
\dot{v} &= f_2(u,v) = - \frac{1}{2 \Omega} \left[ \bar{\gamma} \Omega  v + u \left(-\sigma + \frac{\lambda}{2} \right) - \frac{3}{4} \left( u^2 + v^2 \right) u + \Omega \frac{\bar{\eta}}{4}  \left( u^2 + v^2 \right) v + \bar{F}_0 \cos(\phi) \right], \label{eq:S5}
\end{align}
\end{widetext}
where $\sigma = 1- \Omega^2$.

Also the response of the resonator as a function of the drive frequency has been discussed extensively in Ref.~\cite{Papariello_2016}. The generic situation is that for far red-detuned drive frequencies (for negative Duffing nonlinearity), the system has a unique stable solution with a small amplitude. As the frequency is swept downwards towards resonance, a saddle-node bifurcation occurs at a frequency $\omega^* = \Omega^* \omega_0$ beyond which the system has three solutions, two stable and one unstable solution. This is the frequency at which the response jumps as shown in Fig.~1 in the main text. Further bifurcations occur as the frequency is reduced further. The dimensionless bifurcation frequency $\Omega^*$ can be estimated using a simple bifurcation analysis \cite{Guckenheimer_Holmes}: we first calculate the associated Jacobian matrix $J$ defined as
\begingroup
\renewcommand*{\arraystretch}{1.5}
\begin{equation}
	J(u,v)= \begin{pmatrix}
	\frac{\partial f_1}{\partial u} & 	\frac{\partial f_1}{\partial v}\\
		\frac{\partial f_2}{\partial u} & 	\frac{\partial f_2}{\partial v}
	\end{pmatrix}.
\end{equation}
\endgroup
\hspace{0.1cm}

The Jacobian has two eigenvalues. At a saddle-node bifurcation, one of these eigenvalues becomes zero. The relevant eigenvalue that will show the bifurcation is given by
\begin{widetext}
\begin{align}
\Phi&(\bar{F}_0, \Omega, u^*, v^*) =\frac{-1}{8 \Omega} \left[ 2 \bar{\eta} \Omega \left(u^2 + v^2 \right) + 4 \bar{\gamma} \Omega \right. \nonumber \\
& \left. - \sqrt{(\bar{\eta}^2 \Omega^2 - 27)(u^2 + v^2)^2 + 12 \lambda (u^2 - v^2) + 48 \sigma (u^2 + v^2) + 4 (\lambda^2 -4 \sigma^2) + 8 \lambda \bar{\eta} \Omega u v}\right]_{u\rightarrow u^*, v\rightarrow v^*}, \label{eq:S7}
\end{align}
\end{widetext}
where $u^*$,$v^*$ are the corresponding steady-state solutions of Eq.~\eqref{eq:S4} and~\eqref{eq:S5}. Generally, there are no analytical solutions for $u^*$ and $v^*$ in the nonlinear case. However, for $\Omega \approx \Omega^*$, the resonator is far from resonance, hence the amplitude of the resonator response is very small and essentially dictated by the external drive $\bar{F}_0$. In this regime, the nonlinearities $\alpha$ and $\eta$ can be neglected. In the steady-state, since $\dot{u}=\dot{v}=0$, the solutions to Eq.~\eqref{eq:S5} can therefore be easily obtained
\begin{align}
	\resizebox{0.85\hsize}{!}{$\displaystyle u^*(\Omega, \bar{F}_0) = - \frac{2 \bar{F}_0 (\lambda \cos(\phi)+ 2 \sigma \cos(\phi) + 2 \bar{\gamma} \Omega \sin(\phi))}{\lambda^2 -4 \sigma^2 - 4 \bar{\gamma}^2\Omega^2}\,,$}\\
	\resizebox{0.85\hsize}{!}{$\displaystyle v^*(\Omega, \bar{F}_0)= \frac{2 \bar{F}_0 (\lambda \sin(\phi) - 2 \sigma \sin(\phi) + 2 \bar{\gamma} \Omega \cos(\phi))}{\lambda^2 -4 \sigma^2 - 4 \bar{\gamma}^2\Omega^2}\,.$}
\end{align}
\begin{figure}[t!]
	\includegraphics[width=\columnwidth]{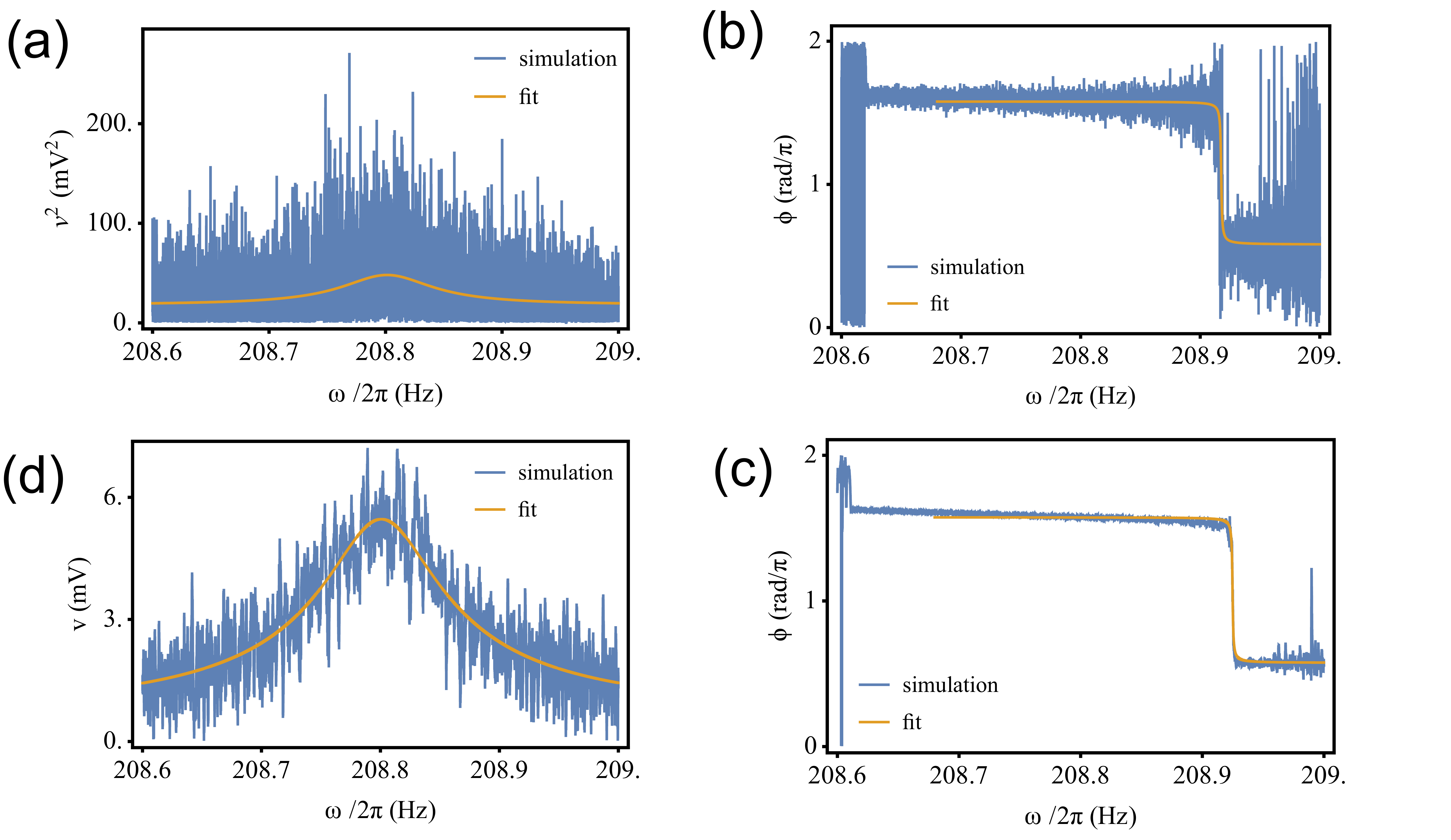}
	\caption{\label{fig:exampleFit} Simulated data and fit used to estimate $V_d$: (a) Squared amplitude and fit of the linear transducer for readout noise. (b) Phase and arctan fit [cf.~Eq.~\eqref{arctanFit}] of the PSBT for readout noise. (c)  Amplitude and Lorentzian fit of the linear transducer for intrinsic noise. (d) Phase and arctan fit [cf.~Eq.~\eqref{arctanFit}] of the PSBT for intrinsic noise. The noise strength is given by $S_{\text{drive}} = 10^{-5} \text{ V}^2/\text{Hz}$ (intrinsic noise) and $\sigma_{\text{readout}} = 3$\, mV (readout noise).}
\end{figure}
Substituting this in Eq.~\eqref{eq:S7}, and fixing all other parameters other than $\bar{F}_0$, the jump frequency $\Omega^*$ is determined by the condition
\begin{equation}
	\Phi(\bar{F}_0, \Omega^*, u^*(\Omega, \bar{F}_0),v^*(\Omega, \bar{F}_0) )=0\,.
\end{equation}
Solving this last condition, we obtain $\Omega^*(\bar{F}_0)$. This equation is a high-order polynomial which cannot be solved analytically. The numerical solutions for the dimensionful $\omega^*$ can then be converted again to physical units of the resonator and our predictions of $\omega^*(\bar{F}_0)$ for different values of the parametric drive amplitude $\lambda$ are plotted in Fig.~1 in the main text. We find that the bifurcation frequency fits the analytical form
\begin{equation}
	\omega^* = \sqrt{a + b \bar{F}_0^2 + \sqrt{c + d \bar{F}_0^2}} + corrections\,,
\end{equation}
where $a$, $b$, $c$ and $d$ are some fitting parameters. For comparison with experimental data we have used the experimentally relevant parameters for $\omega_0$, $\lambda$, $|\alpha|$ and $\eta$ along with a phase offset $\phi = - \pi/10$. We also find that $\omega^*$ depends very weakly on $\eta$.\\

\section{Simulation of the force measurements}

The influence of readout and intrinsic noise on the force measurement can be analyzed by simulating multiple sweeps in the driving frequency $\omega$. For every sweep, one obtains an estimated value for the driving voltage $V_d$. Using multiple sweeps the statistical distribution of these measurements can be obtained (Fig.~3 in the main text). 

The simulations of both the linear transducer and the PSBT use the slow-flow equations (Eq.~\eqref{eq:S4} and \eqref{eq:S5}). Intrinsic noise is simulated by a white-noise process $\xi(t)$ with power spectral-density $S_{\text{drive}}$, while readout noise is taken into account by adding a Gaussian random variable with variance $\sigma_{\text{readout}}^2$ to $u$ and to $v$. 

The system parameters chosen for the simulations are the experimentally-relevant values described above. The driving frequency is continuously swept from $208.6$\,Hz to $209$\,Hz within $10^3$\,s using $10^5$ steps, and we have assumed  that a detection occurs at every 10th step. We used the python package {\it sdeint} and the function {\it itoint} for the numerical integration of the stochastic differential equations for $u$ and $v$. In Fig.~\ref{fig:exampleFit}, an example for sweeps of the linear transducer and the PSBT are shown. In the linear method, the driving voltage $V_d$ is extracted from a Lorentzian fit of the amplitude (squared amplitude) for intrinsic noise (readout noise) while the PSBT uses the function
\begin{equation}
	\phi(\omega) = -\arctan(a(\omega-\omega^*))+b\,,
	\label{arctanFit}
\end{equation}
where $a$ and $b$ are some fitting parameters, to get $\omega^*$ from the phase fit for both types of noise. The driving voltage $V_d$ is then obtained from $\omega^*$. For the linear sensor with intrinsic noise, we used a fit in the squared amplitude because it gives better results (different weighting). By repeating this procedure one obtains many estimates for $V_d$ and the resulting probability distribution function appears in Fig.~3 in the main text.

\end{document}